\documentclass[11pt]{article}
\usepackage{graphics}
\usepackage{latexsym}
\usepackage{algorithmic}
\usepackage{algorithm}
\usepackage{tikz}

\usepackage{tkz-graph}
\usetikzlibrary{shapes}

\usepackage{verbatim}

\newcommand{\bt}{\begin{tabbing}}
\newcommand{\et}{\end{tabbing}}
\newcommand{\PP} {($P_{5}$,$\overline{P}_5$)}
\newcommand{\PPC} {($P_{5}$,$\overline{P}_5$,$C_5$)}

\newtheorem{theorem} {Theorem}

\newtheorem{corollary} {Corollary}

\def\inst#1{$^{#1}$}

\title{Polynomial-time algorithms for minimum weighted colorings of
 ($P_5, \overline{P}_5$)-free graphs and related graph classes}

\author{Ch\'{\i}nh T. Ho\`ang\inst{1}
        \and
        D. Adam Lazzarato\inst{2}
}
\begin{document}
\maketitle

\begin{center}
{\footnotesize

\inst{1} Department of Physics and Computer Science, Wilfrid Laurier University, Waterloo, Ontario, Canada.\\
\texttt{choang@wlu.ca}

\inst{2} Department of Physics and Computer Science, Wilfrid Laurier University, Waterloo, Ontario, Canada.\\
\texttt{adam.lazzarato@gmail.com} }
\end{center}
\begin{abstract}
We design an $O(n^3)$ algorithm to find a minimum weighted
coloring of a ($P_5, \overline{P}_5$)-free graph. Furthermore, the
same technique can be used to solve the same problem for several
classes of graphs, defined by forbidden induced subgraphs, such as
(diamond, co-diamond)-free graphs.

\end{abstract}
\section{Introduction}
Graph coloring is a classical problem in computer science and
discrete mathematics. The chromatic number $\chi(G)$ of a graph
$G$ is the smallest number of colors needed to color the vertices
of $G$ in such a way that no two adjacent vertices receive the
same color. Determining the chromatic number of a graph is a
NP-hard problem. But for many classes of graphs, such as perfect
graphs, the problem can be solved in polynomial time.

Recently, much research have been done on coloring $P_5$-free
graphs. Finding the chromatic number of a $P_5$-free graphs is
NP-hard \cite{KraKra2001}, but for every fixed $k$, the problem of
coloring a graph with $k$ colors admits a polynomial-time
algorithm \cite{HoaKam2008, HoaKam2010}. Research has also been
done on ($P_5, \overline{P}_5$)-free graphs (graphs without $P_5$
and its complement $\overline{P}_5$). In \cite{GiaRus1997}, a
polynomial-time algorithm is found for finding an approximate
weighted coloring of a ($P_5, \overline{P}_5$)-free graph.
Weighted colorings generalize vertex colorings. Given a graph $G$
with a nonnegative integral weight $w_G(v)$ on each vertex $v$ of
$G$, the minimum weighted coloring problem ({\bf MWC}) is to find
stable sets $S_1, S_2, \ldots, S_t$ of $G$ and nonnegative
integers $I(S_1), I(S_2), \ldots , I(S_t )$ such that for each
vertex $v$, $\sum\nolimits_{v\in S_i} I(S_i) \geq w_G(v)$ and that
$\chi_w(G) = \sum\nolimits_{i=1}^{t} I(S_i)$ is as small as
possible; $\chi_w(G)$ is called the {\it weighted chromatic
number} of $G$; the stable sets $S_i$ together with the weights
$I(S_i)$ are called a weighted coloring of $G$.

The motivation of our paper is to find a polynomial-time algorithm
for MWC for \PP -free graphs. In the process of doing this, we
actually solve a more general problem. We prove that for a
hereditary class ${\cal C}$ of graphs, if the minimum weighted
coloring problem can be solved for every prime graph of ${\cal C}$
in polynomial time, then so can the problem for every graph in
${\cal C}$ (definitions not given here will be given later). As a
corollary, we obtain a polynomial-time algorithm to find a minimum
weighted coloring of a ($P_5, \overline{P}_5$)-free graph. This
algorithm runs in $O(n^3)$ time. Furthermore, the same technique
can be used to solve the same problem for several classes of
graphs, defined by forbidden induced subgraphs, such as (diamond,
co-diamond)-free graphs. We will remark on this point in
section~\ref{sec:consequences}. In section~\ref{sec:definition},
we give definitions and discuss the background to our problem. In
section~\ref{sec:algorithms}, we establish the above theorem and
give our algorithm for MWC for \PP -free graphs.

\section{Definitions and background}\label{sec:definition}
Let $G$ be a graph. A set $H$ of vertices of $G$ is a {\it module}
if every vertex in $G-H$ is adjacent to either all vertices of
$H$, or no vertices of $H$; if $|H|=1$ or $|H|=|V(G)|$ then $H$ is
a {\it trivial} module. A graph is {\it prime} if it does not
contain a non-trivial module. For the rest of the paper, modules
are non-trivial unless otherwise noted. A module $H$ is {\it
strong} if for any module $A$, either $H \cap A = \emptyset$, or
$H$ is contained in $A$ or vice versa.
%
%
It is well known (for example, see \cite{McCSpi1994}) that
the vertex set of a graph can be partitioned into unique maximal
strong modules in linear time.

Let $G$ be a graph with a maximal strong module $H$. The graph $G$
can be decomposed into two graphs: one is  $H$ and the other is
the graph $g(G,H,h)$ obtained from $G$ by substituting the vertex
$h$ for $H$, ie. removing $H$ from $G$, adding $h$  and  the edge
$hv$ for every vertex $v \in G-H$ with $vu \in E(G)$ for some $u
\in H$ ($v$ has some neighbor in $H$). If $H$ or $g(G,H,h)$ is not
prime, then we can recursively decompose the graph in the same
way. We can associate this recursive decomposition of $G$ with a
binary tree $T(G)$, where each node $X$ of $T(G)$ represents an
induced subgraph $r(X)$ of $G$, as follows. The root $T$ of $T(G)$
represents $G$ (ie., $r(T)= G$), $T$ has two children $L,R$ where
node $L$ (left child) represents a maximal strong module $H$ and
node $R$ (right child) represents the graph $g(G,H,h)$. If their
representative graphs are not prime, then $L$ and $R$ in turn have
children defined by some maximal strong modules. Thus, the leaves
of $T(G)$ represent prime induced subgraphs of $G$.
Figure~\ref{fig:graphG} shows a graph $G$, Figure
\ref{fig:decompositionTree} shows $T(G)$ together with the
representative graphs of the nodes of $T(G)$. A well known and
easy proof by induction shows that the number of internal nodes of
$T(G)$ is at most $2|V(G)|$ and the total number of edges in all
prime graphs (produced by the decomposition) is at most $|E(G)|$.
There are well known linear time algorithms to construct $T(G)$
and the associative graphs of its internal nodes from $G$
(\cite{CouHab1994,McCSpi1994}, see also the survey paper
\cite{HabPaul2010}).

\begin{figure}
\begin{center}
\begin{tikzpicture}
    \SetVertexNormal[Shape      = circle,
                 LineWidth  = 1pt]
    \Vertex[x=0, y=0]{1}
    \Vertex[x=2 ,y=-2]{3}
    \Vertex[x=2, y=0]{4}
    \Vertex[x=2, y=2]{2}
    \Vertex[x=3.5, y=3]{5}
    \Vertex[x=5, y=1]{6}
    \Vertex[x=5, y=-1]{7}
    \Vertex[x=7, y=2]{8}
    \Vertex[x=9, y=1]{9}
    \Vertex[x=7, y=-2]{10}
    \Vertex[x=9, y=-1]{11}
    \Edge(1)(2)
    \Edge(1)(3)
    \Edge(1)(4)
    \Edge(2)(4)
    \Edge(2)(5)
    \Edge(2)(6)
    \Edge(2)(7)
    \Edge(3)(4)
    \Edge(3)(5)
    \Edge(3)(6)
    \Edge(3)(7)
    \Edge(4)(5)
    \Edge(4)(6)
    \Edge(4)(7)
    \Edge(5)(6)
    \Edge(5)(7)
    \Edge(6)(8)
    \Edge(6)(9)
    \Edge(6)(10)
    \Edge(6)(11)
    \Edge(7)(8)
    \Edge(7)(9)
    \Edge(7)(10)
    \Edge(7)(11)
    \Edge(8)(9)
    \Edge(8)(10)
    \Edge(8)(11)
    \Edge(9)(10)
    \Edge(9)(11)
\end{tikzpicture}
\end{center}
\caption{The graph $G$}\label{fig:graphG}
\end{figure}

\begin{figure}
\begin{tikzpicture}

    \SetVertexNormal[Shape      = ellipse,
                     LineWidth  = 1pt]
    \tikzstyle{every node}=[font=\normalsize]
    \Vertex[x=0, y=6]{G}
    \Vertex[x=-2.5, y=5]{2 3 4}
    \Vertex[L=$g({G,H,h})$,x=2.5, y=5]{rest1} 
    \Vertex[x=-4, y=3]{2 3}
    \Vertex[L=$g({G,H,h})$,x=-1, y=3]{rest2}
    \Vertex[x=1, y=3]{6 7}
    \Vertex[L=$g({G,H,h})$,x=4, y=3]{rest3}
    \Vertex[x=2.5, y=1]{8 9 10 11}
    \Vertex[L=$g({G,H,h})$,x=5.5, y=1]{rest4}
    \Vertex[x=1, y=-1]{10 11}
    \Vertex[L=$g({G,H,h})$,x=4, y=-1]{rest5}
    \Edge(2 3 4)(G)
    \Edge(G)(rest1)
    \Edge(2 3 4)(2 3)
    \Edge(2 3 4)(rest2)
    \Edge(rest1)(rest3)
    \Edge(rest1)(6 7)
    \Edge(rest3)(rest4)
    \Edge(rest3)(8 9 10 11)
    \Edge(8 9 10 11)(rest5)
    \Edge(10 11)(8 9 10 11)

    \draw[fill] (-3.5,5.5) circle [radius=0.03];
    \node [left] at (-3.5,5.5) {2};
    \draw[fill] (-3.5,4.5) circle [radius=0.03];
    \node [left] at (-3.5,4.5) {3};
    \draw[fill] (-3.5,5) circle [radius=0.03];
    \node [left] at (-3.5,5) {4};
    \draw (-3.5, 4.5) -- (-3.5, 5) -- (-3.5, 5.5);

        \draw[fill] (4,5) circle [radius=0.03];
    \node [below] at (4,5) {1};
    \draw[fill] (4.5,5) circle [radius=0.03];
    \node [below] at (4.5,5) {$h_2$};
    \draw[fill] (4.87,5.5) circle [radius=0.03];
    \node [above] at (4.87,5.5) {5};
    \draw[fill] (5.25,5.25) circle [radius=0.03];
    \node [above] at (5.25,5.25) {6};
    \draw[fill] (5.25,4.75) circle [radius=0.03];
    \node [below] at (5.25,4.75) {7};
    \draw[fill] (5.75,5.5) circle [radius=0.03];
    \node [above] at (5.75,5.5) {8};
    \draw[fill] (6.25,5.25) circle [radius=0.03];
    \node [above] at (6.25,5.25) {9};
    \draw[fill] (5.75,4.5) circle [radius=0.03];
    \node [below] at (5.75,4.5) {10};
    \draw[fill] (6.25,4.75) circle [radius=0.03];
    \node [below] at (6.25,4.75) {11};
    \draw (4,5) -- (4.5,5) -- (4.87,5.5);
    \draw (4.87,5.5) -- (5.25,5.25);
    \draw (4.87,5.5) -- (5.25,4.75);
    \draw (4.5,5) -- (5.25,5.25);
    \draw (4.5,5) -- (5.25,4.75);
    \draw (5.25,5.25) -- (5.75,5.5);
    \draw (5.25,5.25) -- (6.25,5.25);
    \draw (5.25,5.25) -- (5.75,4.5);
    \draw (5.25,5.25) -- (6.25,4.75);
    \draw (5.25,4.75) -- (5.75,5.5);
    \draw (5.25,4.75) -- (6.25,5.25);
    \draw (5.25,4.75) -- (5.75,4.5);
    \draw (5.25,4.75) -- (6.25,4.75);
    \draw (5.75,5.5) -- (6.25,5.25);
    \draw (5.75,5.5) -- (5.75,4.5);
    \draw (5.75,5.5) -- (6.25,4.75);
    \draw (6.25,5.25) -- (5.75,4.5);
    \draw (6.25,5.25) -- (6.25,4.75);

    \draw[fill] (-4.75,3.25) circle [radius=0.03];
    \node [above] at (-4.75,3.25) {2};
    \draw[fill] (-4.75,2.75) circle [radius=0.03];
    \node [below] at (-4.75,2.75) {3};

    \draw[fill] (-3,3) circle [radius=0.03];
    \node [above] at (-3,3) {$h_1$};
    \draw[fill] (-2.5,3) circle [radius=0.03];
    \node [above] at (-2.5,3) {4};
    \draw (-3,3) -- (-2.5,3);

    \draw[fill] (1.75,3.25) circle [radius=0.03];
    \node [above] at (1.75,3.25) {6};
    \draw[fill] (1.75,2.75) circle [radius=0.03];
    \node [below] at (1.75,2.75) {7};

    \draw[fill] (5.75,3) circle [radius=0.03];
    \node [below] at (5.75,3) {1};
    \draw[fill] (6.25,3) circle [radius=0.03];
    \node [below] at (6.25,3) {$h_2$};
    \draw[fill] (7,3) circle [radius=0.03];
    \node [below] at (7,3) {$h_3$};
    \draw[fill] (6.62,3.5) circle [radius=0.03];
    \node [above] at (6.62,3.5) {5};
    \draw[fill] (7.5,3.5) circle [radius=0.03];
    \node [above] at (7.5,3.5) {8};
    \draw[fill] (8,3.25) circle [radius=0.03];
    \node [above] at (8,3.25) {9};
    \draw[fill] (7.5,2.5) circle [radius=0.03];
    \node [below] at (7.5,2.5) {10};
    \draw[fill] (8,2.75) circle [radius=0.03];
    \node [below] at (8,2.75) {11};
    \draw (5.75,3) -- (6.25,3) -- (7,3);
    \draw (6.25,3) -- (6.62,3.5);
    \draw (6.62,3.5) -- (7,3);
    \draw (7,3) -- (7.5,3.5);
    \draw (7,3) -- (7.5,2.5);
    \draw (7.5,3.5) -- (8,3.25) -- (8,2.75) -- (7.5,3.5) -- (7.5,2.5) -- (8,3.25);
    \draw (7,3) -- (8,3.25);
    \draw (7,3) -- (8,2.75);

    \draw[fill] (0.5,1.5) circle [radius=0.03];
    \node [above] at (0.5,1.5) {8};
    \draw[fill] (1,1.25) circle [radius=0.03];
    \node [above] at (1,1.25) {9};
    \draw[fill] (0.5,0.5) circle [radius=0.03];
    \node [below] at (0.5,0.5) {10};
    \draw[fill] (1,0.75) circle [radius=0.03];
    \node [below] at (1,0.75) {11};
    \draw (0.5,1.5) -- (1,1.25) -- (1,0.75) -- (0.5,1.5) -- (0.5,0.5) -- (1,1.25);

    \draw[fill] (7.25,1) circle [radius=0.03];
    \node [below] at (7.25,1) {1};
    \draw[fill] (7.75,1) circle [radius=0.03];
    \node [below] at (7.75,1) {$h_2$};
    \draw[fill] (8.5,1) circle [radius=0.03];
    \node [below] at (8.5,1) {$h_3$};
    \draw[fill] (8.12,1.5) circle [radius=0.03];
    \node [above] at (8.12,1.5) {5};
    \draw[fill] (9,1) circle [radius=0.03];
    \node [below] at (9,1) {$h_4$};
    \draw (7.25,1) -- (7.75,1) -- (8.5,1) -- (9,1);
    \draw (7.75,1) -- (8.12,1.5);
    \draw (8.12,1.5) -- (8.5,1);

    \draw[fill] (0,-0.75) circle [radius=0.03];
    \node [above] at (0,-0.75) {11};
    \draw[fill] (-0.5,-1) circle [radius=0.03];
    \node [below] at (-0.5,-1) {10};

    \draw[fill] (5.5,-0.75) circle [radius=0.03];
    \node [above] at (5.5,-0.75) {8};
    \draw[fill] (6,-1) circle [radius=0.03];
    \node [above] at (6,-1) {9};
    \draw[fill] (5.75,-1.5) circle [radius=0.03];
    \node [below] at (5.75,-1.5) {$h_5$};
    \draw (5.75,-1.5) -- (5.5,-0.75) -- (6,-1) -- (5.75,-1.5);
\end{tikzpicture}

\caption{The decomposition tree
$T(G)$}\label{fig:decompositionTree}
\end{figure}

Let $P_k$ (resp., $C_k$) denote the chordless path (resp., cycle)
on $k$ vertices. If $F$ is a set of graphs, then we say a graph
$G$ is $F$-free if $G$ does not contain an induced subgraph
isomorphic to any of the graphs in $F$. A {\it buoy} is the graph
whose vertex set can be partitioned into non-empty sets $S_1, S_2,
S_3, S_4, S_5$ such that there are all edges between $S_i$ and
$S_{i+1}$ and no edges between $S_i$ and $S_{i+2}$ with the
subscript taken modulo 5. A buoy is {\it complete} if every $S_i$
is a complete graph.


Given an ordered graph $(G, <)$, the ordering $<$ is called
\emph{perfect} if for each induced ordered subgraph $(H, <)$ the
greedy algorithm produces an optimal coloring of H. The graphs
admitting a perfect order are called \emph{perfectly orderable}.
%
A stable set of a graph $G$ is \emph{strong} if it meets all
maximal cliques of $G$. (Here, as usual, ``Maximal'' is meant with
respect to set-inclusion, and not size. In particular, a maximal
clique may not be a largest clique.) A graph is \emph{strongly
perfect} if each of its induced subgraphs contains a strong stable
set. In \cite{Chv1984}, it is proved that perfectly orderable
graphs contain strong stable sets and therefore are strongly
perfect.

When $G$ is an input graph to some algorithm, $n(G)$ (resp.,
$m(G)$) denotes the number of vertices (resp., edges) of $G$. When
the context is obvious, we will write $n=n(G)$ and $m=m(G)$.

\begin{theorem}{\rm \cite{Hoa1994}}
\label{thm:strongstable} If there is a polynomial time algorithm A
to find a strong stable set of a strongly perfect graph then there
is a polynomial time algorithm B to find a minimum weighted
coloring and maximum weighted clique of a strongly perfect graph.
If algorithm A runs in time $O(f(n))$ then algorithm B runs in
time $O(n f(n))$. $\Box$
\end{theorem}
In \cite{ChvHoa1987}, it is proved that \PPC -free graphs are
perfectly orderable and that a strong stable set of a \PPC -free
graph can be found in $O(n+m)$ time. So the following result
follows from Theorem~\ref{thm:strongstable}.
\begin{corollary}\label{cor:P5P5barC5}
MWC can be solved for \PPC -free graphs in $O(n(n+m))$ time.
$\Box$
\end{corollary}
In \cite{FouGia1995}, the following result is obtained on the
structure of \PP -free graph with a $C_5$.
\begin{theorem}{\rm \cite{FouGia1995}}\label{thm:buoy}
Let $G$ be a connected \PP -free graph having at least five
vertices. If $G$ contains an induced $C_5$ then every $C_5$ is
contained in a buoy and this buoy is either equal to $G$ or is a
non-trivial module of $G$.  $\Box$
\end{theorem}

\begin{corollary}\label{cor:prime}
A prime \PP -free graph is either $C_5$-free or is the $C_5$.
$\Box$
\end{corollary}
%

In section~\ref{sec:consequences} we will remark on several
classes of graphs and so we need to introduce more definitions
now.

\begin{itemize}
 \item A graph $G$ is {\it chordal} if it does not contain as
 induced subgraphs the chordless cycle $C_k$ for $k \geq 4$.
 \item A graph $G$ is a {\it thin spider} if its vertex set can be partitioned into
a clique $C$ and a stable set $S$ with $|C| = |S|$ or $|C| = |S| +
1$ such that the edges between $C$ and $S$ are a matching and at
most one vertex is not covered by the matching.
 \item  A graph is a {\it thick spider} if it is the complement of a thin spider.
 \item A graph $G$ is {\it matched co-bipartite} if its vertex set can be partitioned into
two cliques $C_1,C_2$ with $|C_1| = |C_2|$ or $|C_1| = |C_2|$ such
that the edges between $C_1$ and $C_2$ are a matching and at most
one vertex is not covered by the matching.
 \item A graph $G$ is {\it co-matched bipartite} if $G$ is the complement of a matched co-bipartite graph.
 \item A bipartite graph $B = (X, Y, E)$ is a {\it bipartite chain
 graph} if there is an ordering $x_1, x_2, \ldots , x_k$ of all vertices
in $X$ such that $N(x_i ) \subseteq N(x_j )$ for all $1 \leq i <
j \leq k$. (Note that then also the neighborhoods of the vertices
from $Y$ are linearly ordered by set inclusion.) If, moreover,
$|X| = |Y| = k$ and $N(x_i ) = \{y_1, \ldots , y_i \}$ for all $1
\leq i \leq k$, then $B$ is prime.
  \item $G$ is a {\it co-bipartite chain graph}
if it is the complement of a bipartite chain graph.
 \item $G$ is an {\it enhanced co-bipartite chain graph} if it can be partitioned into a
co-bipartite chain graph with cliques $C_1, C_2$ and three
additional vertices $a, b, c$ ($a$ and $c$ are optional) such that
$N(a) = C_1 \cup C_2$, $N(b) = C_1$ and $N(c) = C_2$, and there
are no other edges in $G$.
 \item $G$ is an {\it enhanced bipartite chain graph}
if it is the complement of an enhanced co-bipartite chain graph.
\end{itemize}

\section{MWC algorithm for \PP -free graphs}\label{sec:algorithms}
Consider a weighted graph $G$ where each vertex $x$ has a weight
$w_G(x)$. Let $H$ be a non-trivial module of $G$. By $f(G,H,h)$,
we denote the weighted graph obtained from $G$ by substituting a
vertex $h$ for $H$ where the weight function $w$ for $f_w(G,H,h)$
is defined as follows. With $F=f_w(G,H,h)$, for the vertex $h$, we
let $w_F(h) = \chi_w(H)$ and $w_F(x) = w_G(x)$ for all $x \in
G-H$.
\begin{theorem}\label{thm:reduce}
For a weighted graph $G$, we have $\chi_w(f(G,H,h)) = \chi_w(G)$.
Furthermore, given weighted coloring of $f(G,H,h)$  and $H$ with,
respectively, $a$ and $b$ stable sets, a minimum weighted coloring
of $G$ can be constructed in $O(n(a+b))$ time.
\end{theorem}
{\it Proof of Theorem~\ref{thm:reduce}.} Write $F=f(G,H,h)$. We
will first prove $\chi_w(F) \leq \chi_w(G)$. Consider a minimum
weighted coloring of $G$ with stable sets $S_1, S_2, \ldots , S_t$
with each $S_i$ having weight $I(S_i)$. Let ${\cal X}$ be the
stable sets $S_i$ with $S_i \cap H \not= \emptyset$. Write $W =
\sum\nolimits_{S_i \in {\cal X}} I(S_i)$. Since the restriction of
the stable sets of ${\cal X}$ to $H$ is a weighted coloring of
$H$, we have $W \geq \chi_w(H)$. Construct a weighted coloring
$Y_1, Y_2, \ldots$ of $F$ from the stable sets $S_1, S_2, \ldots$
as follows. For each $S_i$, if $S_i \cap H = \emptyset$ then $Y_i
= S_i$; otherwise $Y_i = (S_i - H) \cup \{h\}$. Then let $I(Y_i) =
I(S_i)$. To verify that the stable sets $Y_i$ is a weighted
coloring of $F$, we only need see that $w(h) = \chi_w(H) \leq W =
\sum\nolimits_{y \in Y_{i}} I(Y_i)$. Thus, we have $\chi_w(F)
\leq\sum\nolimits_{i=1}^t I(Y_i) = \sum\nolimits_{i=1}^t I(X_i) =
\chi_w(G)$.

To complete the theorem, we will now prove $\chi_w(F) \geq
\chi_w(G)$. Let ${\cal X}$ (resp., ${\cal Y}$) be the collection
of stable sets $X_1, X_2, \ldots X_a$ (resp., $Y_1, Y_2, \ldots
Y_b$) with weights $I(X_i)$ (resp., $I(Y_i)$) be a minimum
weighted coloring of $H$ (resp., $F=f(G,H,h)$). We can rearrange
the stable sets $Y_i$'s such that there is an integer $c$ such
that $h \in Y_i$ for $i \leq c$, and $h \not\in Y_i$ for $i > c$.
We will describe an algorithm that produces a (minimum) weighted
coloring of $G$ with a collection ${\cal Z}$ of stable sets $Z_i$
and integers $I(Z_i)$ with $\sum\nolimits_{Z_i \in {\cal Z}}
I(Z_i) = \sum\nolimits_{Y_i \in {\cal Y}} I(Y_i) = \chi_w(F)$ (the
detail is spelled out in Algorithm~\ref{alg:merge-color} of the
Appendix). The algorithm takes as input the list ${\cal L}_1$ of
stable sets $X_1, X_2, \ldots X_a$ of $H$, and the list ${\cal
L}_2$ of stable sets $Y_1, Y_2, \ldots Y_b$ of $F$, and produces
the desired sets ${\cal Z}$. We scan sequentially the stable sets
$X_1, X_2, \ldots X_a$ of ${\cal L}_1$ and in parallel the  stable
sets $Y_1, \ldots , Y_c$ of ${\cal L}_2$ and merge them into
stable sets of ${\cal Z}$. Suppose $X_i$ and $Y_j$ are being
scanned. We merge them into a stable set of ${\cal Z}$ by
introduce a stable set $Z_k = X_i \cup Y_j - h$. If $I(X_i) \leq
I(Y_j)$, then we give $Z_k$ the weight of $X_i$, ie. $I(Z_k) =
I(X_i)$, and reduce the weight of $Y_j$ appropriately, ie. $I(Y_j)
= I(Y_j) - I(X_i)$. Now, $X_i$ can be eliminated from the first
list ($Y_j$ remains in the second list if its weight is not zero).
Similarly, if $I(X_i)
> I(Y_j)$, then we give $Z_k$ the weight of $Y_j$, ie. $I(Z_k) =
I(Y_j)$, and reduce the weight of $X_i$ appropriately; now $Y_j$
can be eliminated from the second list. Since
$\sum\nolimits_{i=1}^{c} I(Y_i) \geq \sum\nolimits_{i=1}^{a}
I(X_i) $, after $Y_c$ is processed, all the stable sets in the
first list will be eliminated. Now, the stable sets $Y_{c+1},
\ldots , Y_b$ in the second list are made to be stable sets of
${\cal Z}$; and we have $\sum\nolimits_{Z_i \in {\cal Z}} I(Z_i) =
\sum\nolimits_{Y_i \in {\cal Y}} I(Y_i) = \chi_w(F)$. It is easy
to verify that the stable sets $Z_i$ form a weighted coloring of
$G$. The algorithm produces at most $a+b$ stable sets, and each
stable set has size at most $n$. This establishes the claimed time
bound.
$\Box$

\begin{theorem}\label{thm:prime}
Let ${\cal C}$ be a hereditary class of graphs. If there is an
$O(f(n))$ MWC algorithm  for every prime graph in ${\cal C}$, then
there is an $O(n^2 f(n))$ MWC algorithm for every graph in ${\cal
C}$. $\Box$
\end{theorem}
{\it Proof of Theorem~\ref{thm:prime}.} As remarked in
section~\ref{sec:definition}, the modular decomposition produces
$O(n)$ prime graphs. The result then follows from
Theorem~\ref{thm:reduce}. $\Box$

Now, we turn our attention to solving MWC for weighted \PP -free
graphs.
\begin{theorem}\label{thm:main}
There is a $O(n^3)$ algorithm to solve MWC for $(P_5,
\overline{P_5})$-free graph.
\end{theorem}
{\it Proof of Theorem~\ref{thm:main}.} Let $G$ be a $(P_5,
\overline{P_5})$-free graph. Use the modular decomposition
algorithms of \cite{McCSpi1994} or \cite{CouHab1994} to construct
the decomposition tree $T(G)$ with root $S$. If $G$ is a prime \PP
-free graphs, then $G$ is the $C_5$ or \PPC -free and we are done
by Corollary~\ref{cor:P5P5barC5}. Otherwise, consider the left
child $L$ and the right child $R$ of $S$ in $T(G)$. Let $H$ be the
representative graph of $L$, that is, $r(L)=H$. We know $H$ is a
non-trivial module of $G$. We now recursively solve MWC on $H$ and
$f(G,H,h)$, the latter being the representative graph of $R$.
Given minimum weighted colorings of $H$ and $f(G,H,h)$, we apply
the stable sets merging algorithm of Theorem~\ref{thm:reduce} to
construct a minimum weighted coloring of $G$. The detail is
spelled out in Algorithms~\ref{alg:color}
and~\ref{alg:merge-color} in the Appendix. We start the algorithm
by calling COLOR($S$) on the root $S$ of $T(G)$. We may assume the
total time used by COLOR-PRIME() on all graphs produced by the
algorithm is $O(n(n+m))$ since the total number of edges in all
prime graphs is bounded by $m$. Assume without loss of generality
COLOR-PRIME($G$) returns a minimum weighted coloring of a prime
\PP -graph $G$. An easy proof by induction shows that the number
of stable sets in the minimum weighted coloring produced by the
call COLOR($S$) is at most $2n-1$. Each call to MERGE-COLOR can be
implemented in $O(n^2)$ time. Since the number of internal nodes
of $T(G)$ is $O(n)$ (see \cite{CouHab1994}), the number of calls
to MERGE-COLOR is $O(n)$. It follows our algorithm runs in
$O(n^3)$ time. $\Box$

\section{MWC algorithms for some related graph classes}\label{sec:consequences}
In the previous section, we provide a polynomial time algorithm to
find a minimum weighted coloring of a \PP -free graph. The insight
of our result is that to solve MWC for a hereditary class of
graphs, only prime graphs need to be considered. It turns out that
this idea can be used to solve MWC for several graph classes that
have been studied in the literature. These graph classes are
defined by forbidden certain graphs defined in
Figure~\ref{fig:somegraphs} below. For these classes of graphs, it
has been proved that the prime graphs in the classes have special
structures (such as being perfect) and therefore it is easy to
solve MWC for them. We will now elaborate on this point. Consider
the following theorems.
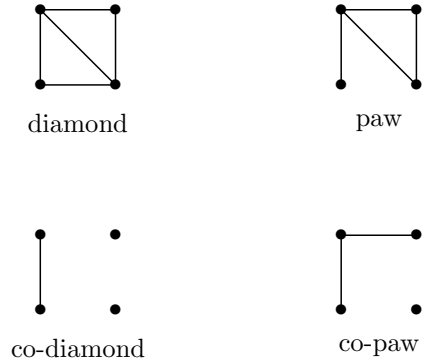
\begin{figure}
\begin{center}
\begin{tikzpicture}
\tikzstyle{every node}=[font=\small]

 \foreach \Point in
{(2,4),(3,4),(6,4),(7,4), (3,0), (3,1), (6,0), (6,1)}{
    \node at \Point {\textbullet};
} \foreach \Point in {(2,3),(3,3),(6,3),(7,3)}{
    \node at \Point {\textbullet};
} \foreach \Point in {(2,1),(7,1)}{
    \node at \Point {\textbullet};
} \foreach \Point in {(2,0),(7,0)}{
    \node at \Point {\textbullet};
}

\draw[line width=0.2mm,black](2,1)--(2,0); \node at (2.5,-0.5)
{co-diamond};

\draw[line width=0.2mm,black](6,0)--(6,1); \draw[line
width=0.2mm,black](6,1)--(7,1); \node at (6.5,-0.5) {co-paw};

\draw[line width=0.2mm,black](2,4)--(3,3); \draw[line
width=0.2mm,black](2,4)--(2,3); \draw[line
width=0.2mm,black](2,4)--(3,4); \draw[line
width=0.2mm,black](3,3)--(3,4); \draw[line
width=0.2mm,black](3,3)--(2,3); \node at (2.5,2.5) {diamond};

\draw[line width=0.2mm,black](6,4)--(7,4); \draw[line
width=0.2mm,black](7,4)--(7,3); \draw[line
width=0.2mm,black](7,3)--(6,4); \draw[line
width=0.2mm,black](6,4)--(6,3); \node at (6.5,2.5) {paw};

\end{tikzpicture}
\end{center}
\caption{Some graphs}\label{fig:somegraphs}
\end{figure}

\begin{theorem}\label{thm:diamond}\cite{BraMah2002}\label{thm:diamondcodiamond}
Let $G$ be a prime graph.\\
\hspace*{2em} (i) If $G$ is (diamond, co-diamond)-free then $G$ or
$\overline{G}$ is a matched co-bipartite graph or G has at most nine vertices.\\
\hspace*{2em} (ii) If $G$ is (paw, co-paw)-free then $G$ is a
$P_4$ or $C_5$.
\end{theorem}

\begin{theorem}\label{thm:p5diamond}\cite{Bra2004} Prime ($P_5$,diamond)-free graphs are either matched
co-bipartite or a thin spider or an enhanced bipartite chain graph
or have at most 9 vertices.
\end{theorem}

There are polynomial time MWC algorithms for all graphs described
in Theorems~\ref{thm:diamond} and~\ref{thm:p5diamond} because
bipartite graphs, co-bipartite graphs, matched co-bipartite graph,
spiders, and enhanced bipartite chain graphs are perfect graphs;
and there is a well known MWC algorithm for perfect graphs
\cite{GroLov1981}.  In some special cases, there are fast MWC
algorithms. For example, spiders are chordal graphs and so the MWC
problem can be solved in $O(n^2)$ time \cite{Hoa1994} on them.

\begin{center}
{\bf Acknowledgement}
\end{center}
This work was partially done as an undergraduate research project
by author D.A.L. under the supervision of author C.T.H., supported
by an NSERC Discovery grant.

\newpage
\begin{center}
{\bf APPENDIX}
\end{center}
\begin{algorithm}
\caption{COLOR($X$) }\label{alg:color}
\begin{algorithmic}
 \STATE{\textbf{input:}} Node $X$ in $T(G)$ with representative
graph $r(X)$, $G$ being a weighted \PP -free graph.
 \STATE{\textbf{output:}} A minimum weighted coloring of $r(X)$.
 \STATE \IF {$X$ is a leaf of $T(G)$}
   \STATE return the output of COLOR-PRIME($r(X)$)
   \ELSE
   \STATE Let $L$ and $R$ be the left and right children of $X$ in
   $T(G)$ where $r(L)$ is a maximal module of $r(X)$
   \STATE Call COLOR($L$) to get a minimum weighted coloring of
   $r(L)$
   \STATE Call COLOR($R$) to get a minimum weighted coloring of
   $r(R)$
   \STATE Call MERGE-COLOR($X,L,R$) and output a minimum weighted
   coloring of $r(X)$
   \ENDIF
\end{algorithmic}
\end{algorithm}
\begin{algorithm}
\caption{MERGE-COLOR($X,L,R$) } \label{alg:merge-color}
\begin{algorithmic}
\STATE{\textbf{input:}}
 \STATE  $X,L,R$ are nodes of $T(G)$
with $L$ (resp., $R$) being the left (resp., right) child of $X$.
 \STATE A minimum weighted coloring of $H=r(L)$ with stable sets
$X_1, X_2, \ldots X_a$ with weights $I(X_i)$
 \STATE A minimum weighted coloring of $f(r(X),H,h)=r(R)$ with stable sets $Y_1, Y_2, \ldots
Y_b$ with weights $I(Y_i)$
 \STATE{\textbf{output:}} A minimum weighted coloring of $r(X)$ with stable sets $Z_1, \ldots, Z_d$ with
weights $I(Z_i)$ with $d \leq a+b$.
 \STATE
  \STATE 1. Enumerate the stable sets of
$f(r(X),H,h)$ as $Y_1, \ldots, Y_c, Y_{c+1}, \ldots Y_{b}$ such
that $h \in Y_i$ if $i \leq c$, and $h \not\in Y_i$ otherwise
 \STATE 2. $i \leftarrow 1, j \leftarrow 1, k \leftarrow 1$
 \STATE 3.
 \WHILE{$i \leq a$}
  \STATE $Z_k \leftarrow X_i \cup Y_j-h$
  \IF {$I(X_i) \leq I(Y_j)$}
    \STATE  $I(Z_k) \leftarrow I(X_i)$
    \STATE $i \leftarrow i+1$
    \STATE $I(Y_j) \leftarrow I(Y_j) - I(X_i)$
    \IF{$I(Y_j) = 0$}
       \STATE $j \leftarrow j + 1$
    \ENDIF
  \ELSE
    \STATE $I(Z_k) \gets I(Y_j)$
    \STATE $I(X_i) \leftarrow I(X_i) - I(Y_j)$
    \STATE $j \leftarrow j+1$
  \ENDIF
  \STATE $k \leftarrow k+1$
\ENDWHILE
 \STATE 4.
 \FOR{$r = j \to b$}
    \STATE $Z_k \gets Y_r$
    \STATE $k \gets k +1$
 \ENDFOR
 \STATE Output the stable sets $Z_1, Z_2 , \ldots $.

\end{algorithmic}
\end{algorithm}


\begin{thebibliography}{99}

\bibitem{Bra2004}
A. Brandst\"adt, ($P_5$,diamond)-free graphs revisited: structure
and linear time optimization, {\sl Discrete Applied Mathematics}
138  (2004),  13--27.

\bibitem{BraHoa2003}
A. Brandst\"adt, C.T. Ho\`ang and V.B. Le,
 Stability number of bull- and chair-free graphs revisited,
 {\sl Discrete Applied Mathematics} 131 (2003), 39--50.

\bibitem{BraLe1999}
A. Brandst\"adt, H.-O. Le and J.-M. Vanherpe,
 Structure and stability number of (Chair, Co-P, Gem)-free graphs,
  {\sl Information Processing Letters} 86 (2003), 161--167.

\bibitem{BraMah2002}
A. Brandst\"adt, S. Mahfud, Maximum Weight Stable Set on graphs
without claw and co-claw (and similar graph classes) can be solved
in linear time, {\sl Information Processing Letters} 84 (2002),
251--259.

\bibitem{BraMos2003}
A. Brandst\"adt, R. Mosca, On variations of $P_4$-sparse graphs,
{\sl Discrete Applied Mathematics} 129 (2003), 521 -- 532.

\bibitem{Chv1984}
    V. Chv\'atal,
    Perfectly orderable graphs,
    {\sl Annals of Discrete Mathematics} 21 (1984), 63-–68.

\bibitem{ChvHoa1987}
   V. Chv\'atal, C. T. Ho\`ang, N. V. R. Mahadev and D. De Werra,
   Four classes of perfectly orderable graphs,
   {\sl J. Graph Theory} 11:4 (1987), 481--495.

\bibitem{CouHab1994}
    A. Cournier and H. Habib,
    A new linear algorithm for Modular Decomposition,
    {\sl Lecture Notes in Computer Science} 787 (1994) 68--84.

\bibitem{FouGia1995}
    J.-L. Fouquet, V. Giakoumakis, F. Maire and H. Thuillier,
    On graphs without $P_5$ and $\overline{P_5}$,
    {\sl Discrete Math.} 146:1-3 (1995) 33--44.

\bibitem{GiaRus1997}
    V. Giakoumakis and I. Rusu,
    Weighted parameters in $(P_5,\overline{P_5})$-free graphs,
    {\sl Discrete Appl. Math} 80 (1997), 255-–261.

\bibitem{GroLov1981}
    M. Gr\"otschel, L. Lov\'asz, and A. Schrijver, The ellipsoid
    method and its consequences in combinatorial optimization,
    {\sl Combinatorica} 1 (1981), 169--197.  (Corrigendum in
    {\sl Combinatorica} 4 (1984), 291--295.

\bibitem{HabPaul2010}
    H. Habib and C. Paul,
    A survey of the algorithmic aspects of modular decomposition,
    {\sl Computer Science Review} 4 (2010) 41--59.

\bibitem{Hoa1994}
    C.T. Ho\`ang,
    Efficient algorithms for minimum weighted colouring of some classes of perfect graphs,
    {\sl Discrete Appl. Math.} 55 (1994) 133–-143.

\bibitem{HoaKam2008}
    C. T. Ho\`{a}ng, M. Kami\'nski, V. Lozin, J. Sawada and X. Shu,
    A note on $k$-colourability of $P_5$-free graphs,
    {\sl Lecture Notes in Computer Science} {5162} (2008)
    387--394.

\bibitem{HoaKam2010}
    C.T. Ho\`ang, M. Kami\'nski, V.V. Lozin, J. Sawada and X. Shu,
    Deciding k-colorability of P5-free graphs in polynomial time,
    {\sl Algorithmica} 57:1  (2010) 74--81.

\bibitem{HoaMaf1992}
    C.T. Ho\`ang, F. Maffray, S. Olariu and M. Preissmann,
     A charming class of perfectly orderable graphs,
     {\sl Discrete Math.} 102 (1992) 67–-74.

\bibitem{KraKra2001}
     J. Kratochvil, D. Kral, Zs. Tuza and G.J. Woeginger,
     Complexity of Coloring Graphs without Forbidden Induced Subgraphs,
     {\sl Lecture Notes in Computer Science} 2204 (2001) 254--262.



\bibitem{McCSpi1994}
    R.M. McConnell and J. Spinrad,
    Modular decomposition and transitive orientation,
    {\sl Discrete Math.} 201 (1999) 189--241.
\end{thebibliography}
\end{document}